\documentclass{ws-procs9x6}
\usepackage{graphicx,epsfig}
\begin{document}

\title{Annihilating dark matter and the galactic positron excess}

\author{Irit Maor}

\address{Institute of Astronomy \\
         University of Cambridge, Madingley Road \\
         Cambridge CB3 0WA, UK \\
         E-mail: i.maor@damtp.cam.ac.uk}
\maketitle

\abstracts{The possibility that the Galactic dark matter is
composed of neutralinos that are just above half the $Z^o$ mass is
examined, in the context of the Galactic positron excess. In
particular, we check if the anomalous bump in the cosmic ray
positron to electron ratio at $10~GeV$ can be explained with the
``decay'' of virtual $Z^o$ bosons produced when the neutralinos
annihilate. We find that the low energy behaviour of our
prediction fits well the existing data. Assuming the neutralinos
annihilate primarily in the distant density concentration in the
Galaxy and allowing combination of older, diffused positrons with
young free-streaming ones, produces a fit which is not
satisfactory on its own but is significantly better than the one
obtained with homogeneous injection.}


\section{Introduction}

The possibility that weakly interacting dark matter particles
(WIMP's) could annihilate into detectable cosmic radiation was
suggested by Silk and Srednicki \cite{Si84}. Tylka and Eichler
\cite{Ty87} noted a reported positron excess, curiously localized
near $10~ GeV$ \cite{Mu85,data,Mu87}, and considered whether it
could be due to the annihilation of photinos (as a simple example
of neutralinos) in the tens of $GeV$ mass range. The difficulty
was that this process, given the laboratory constraints on the
neutralinos, seemed to fall short of providing enough positrons,
and the results were not published. Nevertheless, various papers
on this excess eventually appeared, and noted that the potential
for positron excess could be bolstered by clumpiness in the
annihilating dark matter or by decay of weakly unstable dark
matter particles.

The approach usually found in present literature is to try and fit
the overall $e^+/(e^++e^-)$ ratio, without giving special
attention to the curious behaviour at $10~GeV$
\cite{Hooper:2004bq}. Baltz and Edsjo \cite{Ba01} considered a
whole class of minimal standard supersymmetric models and failed
to get any non-monotonicity in the $e^+/(e^++e^-)$ ratio. They
note that an enhancement factor, presumably due to clumping, of at
least 30 to 100 is needed to account for the observed positron
excess.

Eichler and Maor \cite{p1} considered the possibility that the
observed excess in positrons, and in particular the anomalous
behaviour around $\sim 10~GeV$, is a result of annihilating dark
matter particles with mass just above $m_Z/2$. A non-relativistic
virtual $Z^o$ decay (i.e. when the rest mass of the annihilating
dark matter particle is slightly above 1/2 the $Z^o$ mass)
provides a remarkably good fit to the observed $e^+/(e^++e^-)$
ratio below $10~GeV$. At higher energies, however, the predicted
$e^+/(e^++e^-)$ ratio rises above the observed values, within
conventional assumptions about the injection and propagation. In
particular, it was assumed that the positrons and primary
electrons are each injected homogeneously, with the same spatial
profile, and that their propagation in the Galaxy is identical.
Relaxing the assumption of homogeneous injection for the $Z^o$
decay products was explored in a consequent work \cite{p2}, and
the fit to the data at high energies improved considerably. A
summary of the virtual $Z^o$ decay model is presented here.

\section{The basic equation and its solutions}

The fact that the observed positron excess appears at the
characteristic energy range of positrons from non-relativistic
$Z^o$ decay motivates our consideration of this origin. Such $Z^o$
particles could be produced by the annihilation of neutralinos
that are just beyond half the $Z^o$ mass (so that they do not
contribute to the $Z^o$ decay width). Alternatively, they could be
produced via the slow decay of some other particle that is just
higher than twice the $Z^o$ mass. One could even imagine that such
a particle would couple to non-relativistic matter but not to
relativistic matter, e.g. a massive dilaton coupled to the trace
of $T_{\mu\nu}$, so that it would decay preferentially into $Z$'s,
but not directly into the lighter decay products of the $Z$. \\

In order to achieve enough annihilation of the neutralinos,
clumpiness needs to be invoked. This is a generic problem of
models which make use of dark matter annihilation products. The
fact that clumping is required, and that a likely place for this
is near the Galactic centre, brings us to consider that the
positrons in our neighborhood that are dark matter annihilation
products would have a minimum age. The time needed to diffuse from
the source to our neighborhood, and the age distribution of the
positrons that make it to the Earth's vicinity will result in
fewer young positrons than one associates with the standard leaky
box model. We therefore consider, in addition to standard
assumptions about propagation, that the positrons are injected by
an effectively point source at a finite distance, and look at the
diffused
equation. \\

The steady state diffused equation for the particle number density
is
\begin{eqnarray}
 \frac{\partial n}{\partial t} & = & 0=\hat{D}n-Rn+
        \frac{1}{m_Z}\frac{\partial}{\partial x}\left(m_Z\frac{dx}{dt}n \right)
        +I(x)\delta(r)
\end{eqnarray}
$x=E/m_Z$, $\hat{D}$ is a diffusion operator, $R=B{x}^{0.5}$ is
the escape rate, $m_Z\frac{dx}{dt}=Ax^2$ with $A=8.5\times
10^{-16}~erg/s$ is the Compton loss rate, corresponding to an
electromagnetic energy density in the Galaxy of
$10^{-12}~erg/cm^3$ and $I(x)$ is the injected spectrum.\\

For a homogeneous injection, $\hat{D}n=0$,
\begin{eqnarray}
 n(x) & = & \frac{m_Z}{Ax^2}exp\left[-\frac{2m_ZB}{A\sqrt{x}} \right]
        \int_x^{\infty}I(x')exp\left[
        \frac{2m_ZB}{A\sqrt{x'}} \right]dx'
 \label{hom}
\end{eqnarray}
\\

A one-dimensional diffusion operator with coefficient
${\mathcal{D}}$, $\hat{D}={\mathcal{D}}\frac{\partial^2}{\partial
x^2}$, and with boundary conditions such that $\frac{\partial
n}{\partial r}|_{r=L}=0$ (conserving the number of particles
except for the escape term), gives
\begin{eqnarray}
 n(x,r)  = && \frac{m_Z}{Ax^2}exp\left[-\frac{2m_ZB}{A\sqrt{x}} \right]
        \times \\
   \int_x^{\infty}I(x')exp\left[\frac{2m_ZB}{A\sqrt{x'}} \right]
         \sum_{-\infty}^{\infty} && \cos\left[\frac{\pi n r}{L} \right]
        exp\left[-\left(\frac{\pi n}{L} \right)^2\frac{m_Z{\mathcal{D}}}{A}
        \left(\frac{1}{x}-\frac{1}{x'} \right)\right]dx' \nonumber
 \label{dif}
\end{eqnarray}
While ${\mathcal{D}}$ (the diffusion coefficient) and $L$ (the
size of the leaky box) are free parameters, we took $r=8~Kpc$, the
distance to the galactic centre. $K\equiv {\mathcal{D}}/r^2$ gives
the inverse time for diffusion. \\

\section{Injected spectrums}
\subsection{Backgrounds}
%
%
The injected spectrum of primary electrons is taken to be
$I(x)=Cx^{-2}$.\\
For the positron background we take $Dx^{-2.8}$, presumed to come
from cosmic ray collisions.

\subsection{$Z^o$ Products} We now calculate the spectrum of
positrons resulting from the annihilation of the neutralinos
($\chi)$ via the virtual $Z^o$ channel assuming that $m_{\chi}$ is
just above $m_Z/2$.  It is essentially $Z^o$ decay. We have taken
three generations of particle families, except for the top quark
since it is much heavier then the mass of the $Z^o$. All other
particles were considered massless, and the calculation is in
zeroth order. The positrons' as well as the electrons' injected
spectrum has contributions from the following channels:
\begin{itemize}

\item
{\bf{$Z\rightarrow e\bar{e}$}}\\
The branching ratio of direct decay to electron-positron pair is
$\frac{\Gamma(Z\rightarrow e\bar{e})}{\Gamma(Z\rightarrow
all)}=0.0344$. The energy spectrum of these positrons is
$\delta(x-\frac{1}{2})$.

\item
{\bf{$Z\rightarrow \mu\bar{\mu}$}}\\
All the produced muons eventually decay into $\nu^{\mu}\nu^e e$,
in an exchange of a W boson. The resulting spectrum energy
\cite{Halzen:mc} in the muon rest frame is:
\begin{eqnarray}
   I(\epsilon)=\frac{16}{m_{\mu}^4}
        \left(3m_{\mu}\epsilon^2-\epsilon^4\right),~~~~~~~
        0<\epsilon<\frac{m_{\mu}}{2} \nonumber
\end{eqnarray}
where $\epsilon$ is the positron's energy in the muon's rest
frame. To boost the spectrum into the observer's frame, we have
taken $\beta\approx 1$ and $\gamma=m_Z/(2m_{\mu})$. Assuming the
muon decay is isotropic:
\begin{eqnarray}
   I(x)=\frac{2}{3}\left[5-36x^2+32x^3 \right]  \nonumber
\end{eqnarray}
with a similar branching ratio, $0.0344$.

\item
{\bf{$Z^o\rightarrow \tau\bar{\tau}$}}\\
The tau can decay into 3 colors of $ud$ pairs, into electrons
directly, or into muons which then decay into electrons. For the
20\% of taus which decay directly into electrons, the calculation
is the same as the above, with a branching ratio of
$0.0344\times 0.2$.\\

For the 20\% of the $\tau\rightarrow\mu\rightarrow e$ channel, the
resulting contribution is:
\begin{eqnarray}
 I(x)=\frac{2}{9}\left[-\frac{95}{3}-108x^2+\frac{1408}{3}x^3
 -\left(25+324x^2+128x^3\right)\ln \left(2x\right) \right]  \nonumber
\end{eqnarray}
with a branching ratio of $0.0344\times 0.2$.\\
We have neglected the 60\% of the $\tau\rightarrow q$.

\item
{\bf{$Z\rightarrow q\bar{q}$}} \\
The hadronic channel of the $Z$ decay eventually contributes some
electrons and positrons through the production of pions. We have
estimated the energy spectrum of the pion production from
\cite{had}:
\begin{eqnarray}
    P(E_{\pi})&=&10^{a_k-b_k\bar{x}} \nonumber \\
    a_1=3,~b_1=10~~~&&~~~0<\bar{x}<0.1 \nonumber \\
    a_2=2,~b_2=4 ~~~&&~~~0.1<\bar{x}<1 \nonumber
\end{eqnarray}
with $\bar{x}\equiv 2E_{\pi}/m_Z$. \\
In the pion's rest frame, the energy spectrum of the electrons is
close to a delta function, with an energy of $\sim 45~MeV$. In
transferring back to the lab frame, we have used $\beta\approx 1$
and $\gamma=E_{\pi}/m_{\pi}$. The resulting contribution is:

\begin{eqnarray}
    I_{h}(x)=\frac{14}{9}\int_{\frac{28}{9}x}^1 \frac{d\bar{x}}{\bar{x}}
        10^{a_k-b_k\bar{x}} \nonumber
\end{eqnarray}
The branching ratio of of the $Z^o$ decay into hadrons is
$0.6916$.
\end{itemize}

After taking the right weight of each of these channels, the
injected spectrums are:
\begin{eqnarray}
 && I(x) =  0.6916NI_h(x)+0.0344N\times  \label{is} \\
        && \left[\delta\left(x-\frac{1}{2}\right)+\frac{70}{27} -
        \frac{168}{5}x^2+\frac{6272}{135}x^3-\left(\frac{10}{9}+
        \frac{72}{5}x^2+\frac{256}{45}x^3\right)
        \ln \left(2x\right) \right]  \nonumber
\end{eqnarray}
where $N$ is the annihilation rate density. The needed injection
rate to supply the observed positrons is about $1 \times
10^{-29}cm^{-3} s^{-1}$ above $10~ GeV$. The actual annihilation
rate density N therefore needs to be about $(2.2 \times
0.034)^{-1}$ times that, $N=1.3\times 10^{-29}cm^{-3} s^{-1}$.
Here the factor of $2.2$ assumes that $Z$ decay into electrons,
muons, and $20$ percent of the taus results in positrons above $10
~GeV$. The need for significant dark matter clumping has not been
avoided in the present scenario.

\section{Results}

\begin{figure}
 \centerline{
 \epsfxsize=7cm
 \epsfbox{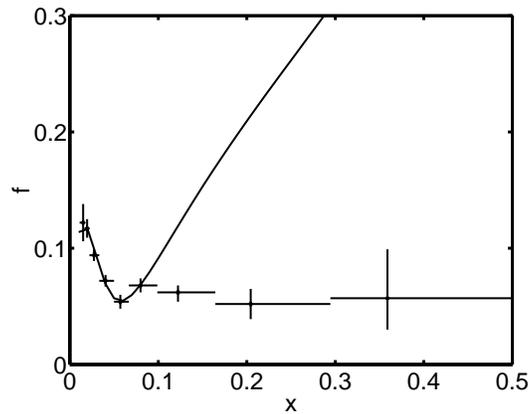}}
 \caption{
    $f\equiv e^+/(e^++e^-)$ as a function of $x\equiv E/m_Z$.
    Our model compared with HEAT 94+95 data 
    The data and its error bars are marked with the crosses.
    The parameters of this example are $A=8.5\times
    10^{-16}~erg/s$, $B=7.6\times 10^{-15}~1/s$,
    $C=9.1\times 10^{-28}~e/(cm^3\cdot s)$, and
    $D=1.3\times 10^{-31}~p/(cm^3\cdot s)$.
 \label{p1}}
\end{figure}

We first consider a homogeneous injection of both backgrounds and
the $Z^o$ products, see Figure~\ref{p1}. A striking feature of
this figure is the good fit below $10~GeV$, which is mainly due to
the positrons that emerge from decaying muons. Above $10~GeV$,
however, the fit grows very quickly, which is not compatible with
the plateau that the data suggests. The reason for this rise is
that some $Z^o$'s decay directly into high energy $e^+e^-$ pairs
so that the $e^+$ energy is half the $Z^o$ mass, and this gives
rise to a high energy bump in the $e^+/(e^++e^-)$ ratio at about
$50~GeV$. Even though the branching ratio into direct $e^+e^-$ is
rather small (0.0344), it is dominant over the power-law
background contribution at this energy. While this bump can be
partially washed out by losses and escape, it was found that the
high energy $e^+/(e^++e^-)$ ratio is nevertheless apparently too
high to fit the observations. As discussed in Eichler (1989) this
is a generic problem for any positron source that is significantly
harder than the primary electrons above $10~GeV$.

\begin{figure}
 \centerline{
 \epsfxsize=7cm
 \epsfbox{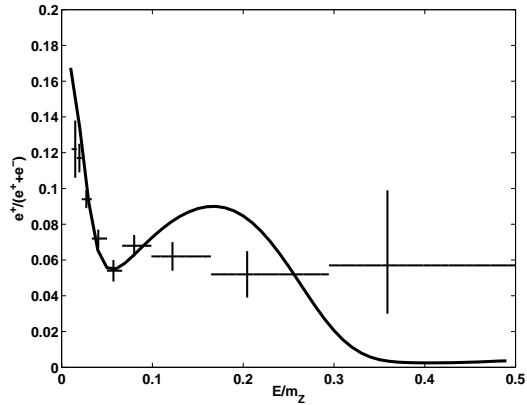}}
 \caption{The $e^+/(e^++e^-)$ as a function of $x=E/m_Z$, for a
            single source $Z$ decay injected spectrum.
            $A=8.5\times 10^{-16}~\frac{erg}{s}$,
            $B=7.1\times 10^{-15}~\frac{1}{s}$,
            $C=4.0\times 10^{-29}~\frac{1}{cm^3~s}$,
            $D=1.3\times 10^{-31}~\frac{1}{cm^3~s}$,
            and $K=1.9\times 10^{-16}~\frac{1}{s}$.
 \label{singleZ}}
\end{figure}

Next we consider an inhomogeneous injection for the $Z^o$
products, while still keeping the backgrounds homogeneous.
Figure~\ref{singleZ} shows a fit with a single point source of
$Z^o$ decay and a single diffusion coefficient. The good fit to
the low energies from Figure~\ref{p1} is still present, but the
excess in energies toward $x=1/2$ is now is suppressed by the
finite age effect. As the figure shows, we are now facing a
scenario which is opposite to the homogeneous injection case; the
finite age effect tends to suppress the high energy excess at the
price of killing it off altogether.

\begin{figure}
 \centerline{
 \epsfxsize=7cm
 \epsfbox{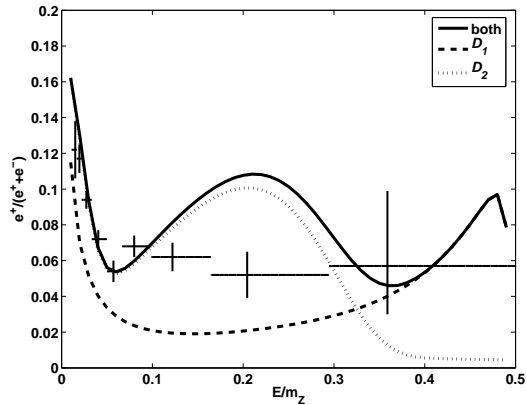}}
 \caption{The $e^+/(e^++e^-)$ as a function of $x=E/m_Z$, for a
            combination of 2 sources of $Z$ decay injected spectrum.
            $A=8.5\times 10^{-16}~\frac{erg}{s}$,
            $B=7.6\times 10^{-15}~\frac{1}{s}$,
            $C=4.9\times 10^{-29}~\frac{1}{cm^3~s}$,
            $D=1.3\times 10^{-31}~\frac{1}{cm^3~s}$,
            $K_1=2.8\times 10^{-14}~\frac{1}{s}$
            and $K_2=2.8\times 10^{-16}~\frac{1}{s}$. The
            ratio between the two components is $1:5$.
%
 \label{combo}}
\end{figure}

However,  there are several possibilities that dispel this
problem:  There may be more than one source, and there may be more
than one route by which the particles diffuse  or free-stream from
the source to our vicinity. High energy particles diffuse much
less than low energy ones because they are fewer in number and
create less waves. So their self-generated scattering is less
efficient. Thus, the fraction of free streaming particles should
be higher at higher energy. Figure~\ref{combo} shows a combination
of two $Z^o$ decay components, an older, larger one that arrives
via diffusion, and a younger, smaller component that has managed
more free streaming. This figure illustrates that if one takes an
age distribution into account, the flexibility in adjusting the
high energy spectrum becomes much larger, and can be fitted to the
data.

\begin{figure}
 \centerline{
 \epsfxsize=7cm
 \epsfbox{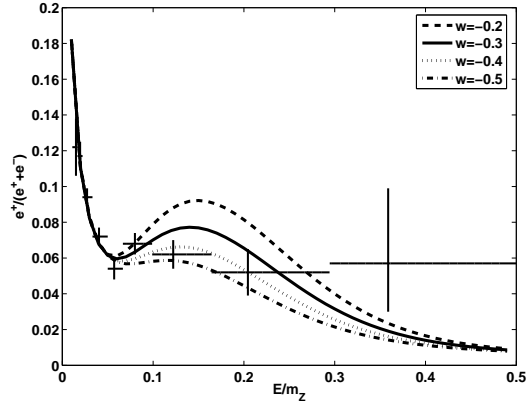}}
 \caption{The $e^+/(e^++e^-)$ as a function of $x=E/m_Z$, for
            various power laws, $Nx^{w}$, as the injected spectrum.
            $A=8.5\times 10^{-16}~\frac{erg}{s}$,
            $B=4.4\times 10^{-15}~\frac{1}{s}$,
            $C=1.7\times 10^{-29}~\frac{1}{cm^3~s}$,
            $D=1.1\times 10^{-31}~\frac{1}{cm^3~s}$,
            and $K=6.6\times 10^{-17}~\frac{1}{s}$.
 \label{mpl}}
\end{figure}

\begin{figure}
 \centerline{
 \epsfxsize=7cm
 \epsfbox{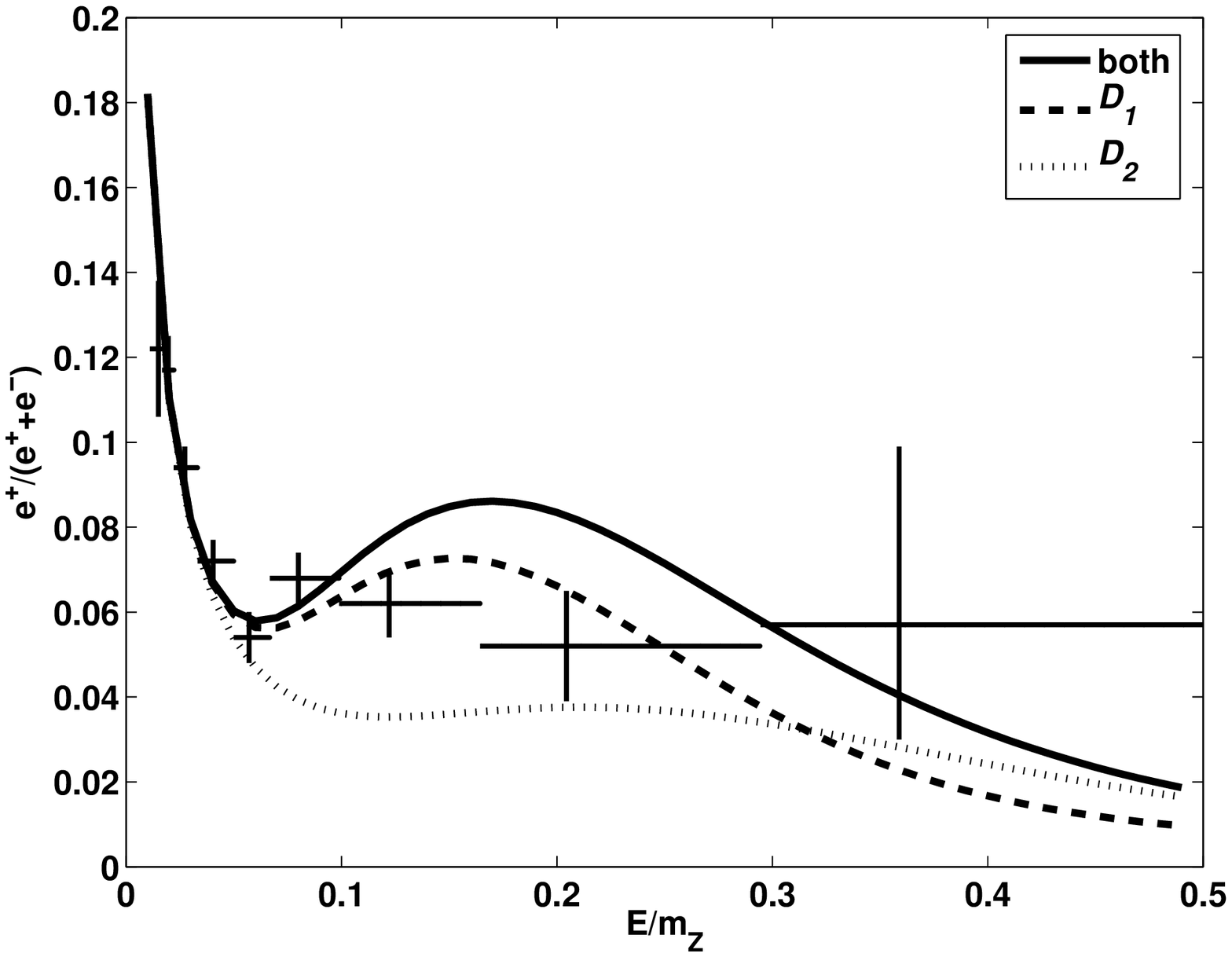}}
 \caption{The $e^+/(e^++e^-)$ as a function of $x=E/m_Z$, for a
            combination of 2 sources of power law ($w=-0.3$)
            injected spectrum.
            $A=8.5\times 10^{-16}~\frac{erg}{s}$,
            $B=4.4\times 10^{-15}~\frac{1}{s}$,
            $C=1.7\times 10^{-29}~\frac{1}{cm^3~s}$,
            $D=1.1\times 10^{-31}~\frac{1}{cm^3~s}$,
            $K_1=7.2\times 10^{-17}~\frac{1}{s}$
            and $K_2=1.2\times 10^{-16}~\frac{1}{s}$. The
            ratio between the two components is $10:1$.
 \label{plcombo}}
\end{figure}

For sake of comparison, we also include a power law injected
spectrum. Figure~\ref{mpl} shows various power laws, and
Figure~\ref{plcombo} shows two components with different ages. We
find that as long as the injected power law is hard enough, one
can produce a low energy dip. While we find that the goodness of
fit provided by $Z^o$ decay and by power law are qualitatively the
same, we have deliberately not quantified this with the standard
statistical measures. The $GeV$ dip which we are trying to address
formally has less statistical weight compared with the overall
shape of the data, but it is the more striking observational
result and the one that seems hardest to explain. Trying to get
the best parameters (for either power law or $Z^o$ decay as
injected spectrums) would wash out the low energy behavior that we
are focusing on. As we are interested in exploring the possibility
of connecting the $GeV$ behavior to properties of dark matter more
than constraining parameters, we settle for the time being in
showing characteristic examples.

\section{Discussion}

We have considered a  scenario for positron excess via
non-relativistic $Z^o$ decay in which the energies of the decay
products are as low as possible and the neutralino annihilation
rate that would give rise to the $Z^o$'s as high as possible. We
considered homogeneous injection, as well as injection from a
point source at a finite distance.

Homogeneous injection fits the data remarkably well for energies
below $10~GeV$, but produces too many high energy positrons
(originating from the direct $Z^o \rightarrow e^+e^-$ channel). A
single source with a single diffusion coefficient does not provide
a good fit to the high energy either, and the most promising
scenario is to allow a small fraction of the positrons to free
stream and arrive at the Earth's vicinity much younger than the
rest.

Although we can reproduce the $7~GeV$ dip, the peak at $E\sim 15~
GeV$ is still too big for the HEAT data. This seems to be a
generic feature of our results, regardless of whether the
injection source is virtual $Z^o$ decay or a power law. The
problem would be worse if the virtual $Z^o$ had an energy well
above $m_Z$. Knowing that the virtual $Z^o$ must be close to its
mass shell if it is to provide a decent fit suggests that its loop
corrections would be large and it might be discernable or
falsifiable with particle collider data. Further possible
observational consequences will be considered in a forthcoming
publication \cite{p2}.


\section*{Acknowledgments}
The work described here was done in collaboration with David
Eichler. My sincere thanks to him, for an enjoyable and
interesting collaboration. I gratefully acknowledge the support
from the Leverhulme Trust.

\vspace{2cm}

\end{document}